\documentclass[aps, prb, 10pt, twocolumn, notitlepage, superscriptaddress]{revtex4-2}
\usepackage[colorlinks=true,allcolors=red!80!black]{hyperref}
\usepackage{amssymb}
\usepackage{amsmath}
\usepackage{graphicx}
\usepackage{subfigure}
\usepackage{amsfonts}
\usepackage{xcolor}
\usepackage{epsfig}
\usepackage{color}
\usepackage{bbm}
\usepackage{bm}
\usepackage{tabularx}
\usepackage{multirow}
\usepackage{mathtools}
\usepackage{xfrac}
\usepackage{mathptmx}
\DeclareMathAlphabet{\mathcal}{OMS}{cmsy}{m}{n}

\newcommand{\ket}[1]{\vert{#1}\rangle} 
\newcommand{\bra}[1]{\langle{#1}\vert} 
 
\newcommand{\trans}[2]{\ket{#1}\!\bra{#2}}
\newcommand{\mean}[1]{\langle #1 \rangle}

\newcommand{\one}{\openone}
\newcommand{\abs}[1]{\left|#1\right|}

\newcommand{\half}{\tfrac{1}{2}}

\newcommand{\vnnb}{V$_{\rm N}$N$_{\rm B}$}

\makeatletter
\newcommand*\bigcdot{\mathpalette\bigcdot@{.5}}
\newcommand*\bigcdot@[2]{\mathbin{\vcenter{\hbox{\scalebox{#2}{$\m@th#1\bullet$}}}}}
\makeatother

\newcommand{\liov}{\mathcal{L}}
\newcommand{\diss}{\mathcal{D}}
\newcommand{\proj}{\mathcal{P}}
\newcommand{\comp}{\mathcal{Q}}

\newcommand{\vb}{V$_{\rm B}$}

\begin{document}
\title{Strain induced coupling and quantum information processing with hexagonal boron nitride quantum emitters}

\author{F. T. Tabesh}
\affiliation{Department of Physics, Isfahan University of Technology, Isfahan 84156-83111, Iran}

\author{Q. Hassanzada}
\affiliation{Fritz-Haber-Institut der Max-Planck-Gesellschaft, Faradayweg 4-6, D-14195 Berlin, Germany}
\affiliation{Department of Physics, Isfahan University of Technology, Isfahan 84156-83111, Iran}

\author{M. Hadian}
\affiliation{Department of Physics, Isfahan University of Technology, Isfahan 84156-83111, Iran}

\author{A. Hashemi}
\affiliation{Department of Applied Physics, Aalto University, P.O. Box 11100, 00076 Aalto, Finland}

\author{I. Abdolhosseini~Sarsari}
\affiliation{Department of Physics, Isfahan University of Technology, Isfahan 84156-83111, Iran}

\author{M. Abdi}
\affiliation{Department of Physics, Isfahan University of Technology, Isfahan 84156-83111, Iran}

\begin{abstract}
We propose an electromechanical scheme where the electronic degrees of freedom of boron vacancy color centers hosted by a hexagonal boron nitride nanoribbon are coupled for quantum information processing.
The mutual coupling of color centers is provided via their coupling to the mechanical motion of the ribbon, which in turn stems from the local strain. The coupling strengths are computed by performing \textit{ab initio} calculations. The density functional theory (DFT) results for boron vacancy centers on boron nitride monolayers reveal a huge strain susceptibility.
In our analysis, we take into account the effect of all flexural modes and show that despite the thermal noise introduced through the vibrations one can achieve steady-state entanglement between two and more number of qubits that survives even at room temperature. Moreover, the entanglement is robust against mis-positioning of the color centers.
The effective coupling of color centers is engineered by positioning them in the proper positions. Hence, one is able to tailor stationary graph states.
Furthermore, we study the quantum simulation of the Dicke-Ising model and show that the phonon non-equilibrium phase transition occurs even for a finite number of color centers.
Given the steady-state nature of the proposed scheme and accessibility of the electronic states through optical fields, our work paves the way for the realization of steady-state quantum information processing with color centers in hexagonal boron nitride membranes.
\end{abstract}

\maketitle

%
%
\section{Introduction}
The discovery of color centers in hexagonal boron nitride (hBN) monolayers and nanotubes counts as a big step forward in the nanophotonics where miniaturization of the devices, yet enhancement of their performance is the goal~\cite{Tran2016a, Chejanovsky2017}.
These point defect color centers were later shown to be beneficial in the field of optomechanics and the related quantum technologies~\cite{Abdi2017, Abdi2019, Abdi2021}. The considerably low mass and high quality mechanical properties of hBN structures makes them promising as quantum objects with motional degree of freedom in optomechanical systems~\cite{Atalaya2008, Song2010, Golberg2010}.
An experimentally feasible approach for reliable coupling of the quantum emitters to the motion requires a high sensitivity of the point defect structure responsible for the emissions to the local strain. The primary experiments reporting on the strain effect in hBN emitters where based on bulk samples upon a substrate and hence showed small susceptibilities~\cite{Grosso2017}.
In the later experiments larger effect of the strain was further revealed by modifying the setup~\cite{Yim2020, Mendelson2020}.
On the theory side, the computational analysis on the effect of strain on the antisite complex defects \vnnb, that are counted as one of the main candidates of quantum emission from hBN, proved that the sensitivity of these color centers to the local strain is significant~\cite{Li2020}.
Meanwhile, there are other point defects responsible for the single-photon emission such as boron vacancy that has shown interesting electronic and magnetic properties such as optical polarization of their ground state~\cite{Abdi2018, Gottscholl2020}.
Notwithstanding, a strain study on its emission spectrum is still lacking.

The advantages of color centers are encouraging for the realization of quantum interfaces with various quantum systems, including mechanical vibrations. The most straightforward approach for having an electronic-motion interface is to couple them through local strain, if the sensitivity of color centers to the locally applied strain allows for it. This has already been investigated and proven rather successful in diamond point defects~\cite{Bennett2013, Teissier2014, Ovartchaiyapong2014}, in which, some schemes are even proposed for entangling and networking the color centers or nuclear spins through the phonons~\cite{Albrecht2013, Lemonde2018, Cao2018}.
In particular, the exceptional strain susceptibility of silicon vacancy centers in diamond has allowed the researchers to develop setups where a considerable number of SiV centers are controlled and tuned by strain~\cite{Sohn2018, Meesala2018, Maity2020}.
On the other hand, the promising mechanical properties of carbon nanotubes have led the researchers to an alternative direction where the excitons play the role of a quantum interface~\cite{WilsonRae2012, Hofmann2013, Ma2015}.
In these two directions, while the former lacks low-mass mechanical resonators, and thus smaller coupling rates, the latter suffers limited accessibility of the excitons.

Here, we propose a scheme with experimental feasibility where the advantages of the two above mentioned systems are gathered in the same device. Namely, the accessible color centers with high strain susceptibilities are brought into the low mass, high quality mechanical resonators at nanoscales.
Our device benefits from the coupling of color centers to the flexural modes of a freestanding boron nitride nanoribbon (BNNR) through the local strain.
By performing \textit{ab initio} computations for the uniaxial and biaxial strain on the negatively charged boron vacancy defects on hBN monolayer we find a large shift in the zero-phonon line emissions when the strain is applied along the armchair direction.
This signals huge strain susceptibility of this kind of defect in freestanding membrane structures supported or clamped along their zigzag edges, see Fig.~\ref{fig:scheme}.
Using these promising results we find the strength of coupling between the electronic levels of quantum emitters to the flexural modes of a BNNR resonator.
The setup is shown to be efficient for the realization of a qubit-qubit coupling between the electronic states of distant color centers.
We show that by engineering the system geometry the strength of mutual coupling becomes comparable to the decay rate of emitters. As an example of employing the created interaction, we study the possibility of generating steady-state entanglement in the system consisting of two and more color centers.
The feasibility of the proposed entangling mechanism is corroborated by our numerical simulations, where realistic sources of noise, especially vibrational thermal noise, are accounted for.
We also show that graph states with customizable forms are prepared as resources for the quantum metrology~\cite{Shettell2020}.
Due to their long-living nature, we believe that such steady-state entanglements can be the first quantum technological achievements of hBN color centers with application in quantum enhanced sensing.
From a different point of view, our scheme can serve as a quantum simulator for spin-spin and spin-boson systems. To show its capabilities, in this work we study the phase transition of a Dicke-Ising model and show that our scheme exhibits the phonon superradiance even with a finite number of color centers. We then investigate the effect of coherent and incoherent inter-qubit interactions on the phase transition.

The paper is organized as follows: in the next section we present the DFT calculations and put forth a Hamiltonian that describes the system dynamics.
In Sec. III we derive a reduced master equation for color centers by eliminating the vibrational degrees of freedom.
The study on steady-state entanglement among two and several color centers are presented in Sec. IV and V, respectively.
The phonon driven-dissipative superradiance phase transition the effect of inter-qubit interactions is investigated in Sec. VI.
The concluding remarks and a summary is given in Sec. VII.

%
%
\section{Model}
To describe our scheme, we first provide DFT computational evidence for the large frequency shift imposed on the negatively charged \vb\ color centers on a monolayer of hBN when they are under local strain along the armchair direction.
Based on this, the Hamiltonian of a system of BNNR with embedded quantum emitters is given. We then specify details of the proposed device.

\subsection{Electronic structure study}
We utilize density functional theory and constrained occupation density functional theory (CDFT) as implemented in the \textsc{vasp} code~\cite{Kresse1996} to study the effect of strain on electronic structure and ZPL energy of negatively charged boron vacancy defect in hBN monolayer.
We employ an energy cutoff of 450 eV for the plane wave basis set within the projector augmented-wave method (PAW)~\cite{Blochl1994, Bengone2000} as the pseudo-potential treatment. 
The failure of DFT on bandgap calculation led us to employ Heyd, Scuseria, and Ernzerhof functional (HSE)~\cite{Heyd2003, Krukau2006} method with a mixing parameter of $\alpha=0.41$ to attain a band gap of 6.4 eV with counting zero-point renormalization (ZPR) correction due to electron-phonon coupling~\cite{Tutchton2018,Turiansky2019}.

An orthorhombic supercell, including 55 Boron atoms and 56 Nitrogen atoms is constructed to provide zigzag and armchair directions along $a$ and $b$ axes, respectively.
The vacuum size is set to 15~\AA\ to separate the periodic images. A single k-point at the Gamma point is used for sampling the Brillouin zone. We have performed the calculations for one data point with a larger supercell for checking the validity of our computations, and the results show that a single k-point is sufficient for the supercell of properties described above. The energy and force tolerances for structure optimization are considered to be $10^{-4}$ eV and $0.01$ eV/\AA, respectively.
Within the CDFT method, the ions are relaxed to reach the global minimum of the adiabatic potential energy surface in the excited state. In this method, one electron is promoted to a higher electronic level, and a hole remains in its original level. The ZPL energy is then obtained by computing the difference in total energy between the two electronic states (ground state and excited state) minima.

\begin{figure*}[tb]
\includegraphics[width=\textwidth]{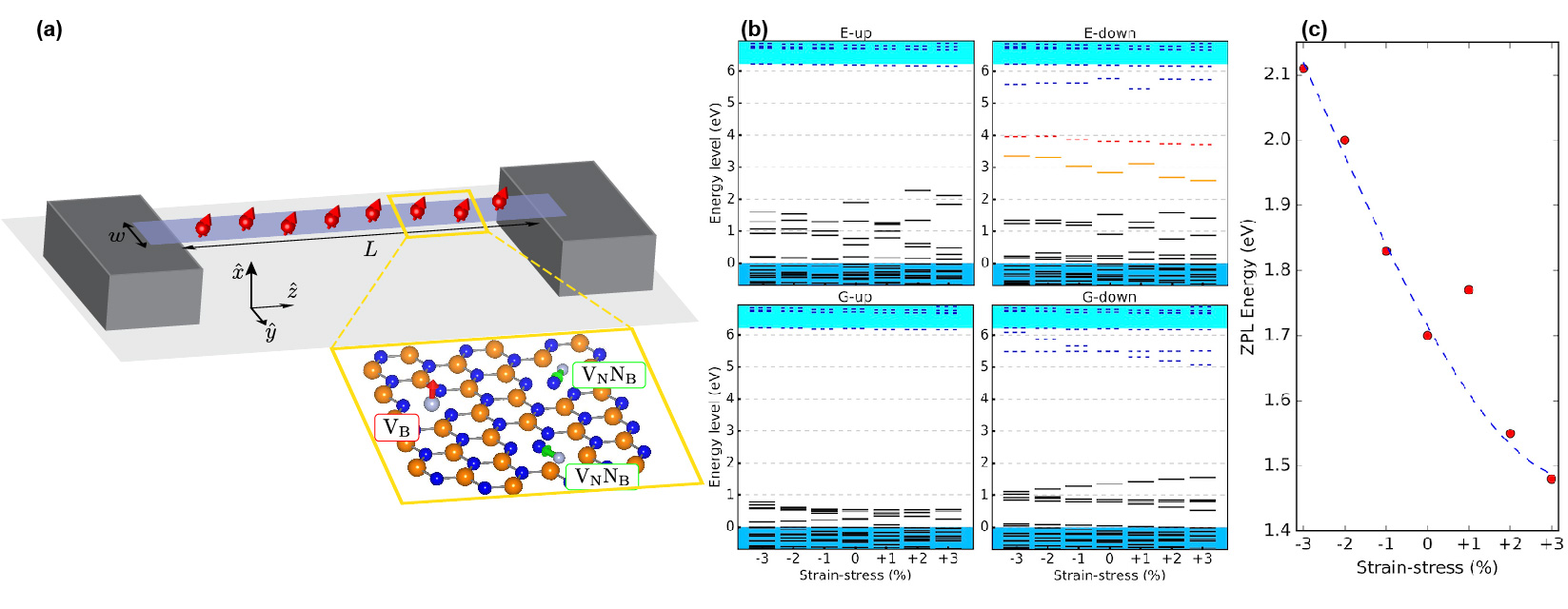}
\caption{%
(a) Sketch for the proposed setup: The boron nitride nano-ribbon with embedded color centers. The inset shows geometry of the hBN membrane with \vb\ and \vnnb\ embedded color centers. The main dipole moments are shown as red and green arrows.
(b) The ground state and excited state electronic structure under uniaxial strain along the armchair direction. The defect excited state and ground state energy levels in spin-down and spin-up channels are denoted as E-down, E-up, G-down, and G-up, respectively. The solid black lines and blue dashed lines show the occupied and unoccupied states. Red dashed lines and solid orange lines display initial and final states during the excitation.
(c) Variation of the zero-phonon line with the applied strain; the red spots show the extracted data and the blue dashed line is a numerical fit.
}%
\label{fig:scheme}%
\end{figure*}

The Poisson ratio of the hBN monolayer has been predicted to be 0.211~\cite{Boldrin2011}.
By considering this value, we apply uniaxial and biaxial in-plane stress-strain on \vb\ defect in the range of $-3\%$ to $+3\%$ in $1\%$ increment.
The results for the ZPL energy in Fig.~\ref{fig:scheme} show that an increasing uniaxial strain along the armchair direction almost linearly decreases the ZPL energy of $^{3}\!\textrm{E}'\rightarrow\ ^{3}\!\textrm{A}'_{2}$ transition in the \vb\ defect from 2.1 to 1.5 eV~\cite{Abdi2018}.
We, however, notice an abnormal local maximum around $+1\%$, which is attributed to the complicated Jahn-Teller nature of \vb. Full understanding of this feature is out of scope of this work and subject of a future study.
For the completeness, the results on the effect of uniaxial strain along zigzag direction as well as biaxial strain are reported in Appendix~\ref{sec:DFTstudy}.

%

\subsection{Hamiltonian of the system}
The system is composed of a freestanding membrane of hBN with embedded color centers.
The flexural vibrations of the membrane can create local strains at the point defects leading to a coupling between electronic states of the color centers and the motional modes via a deformation in the molecular orbitals of the color centers.
The \textit{ab initio} analysis provided in the previous section has allowed us to find the effect of local strain on the energy levels of negatively charged boron vacancy defects (V$_{\rm B}$). Our results suggest that the deformation potential for the simulated configuration (a monolayer hBN) becomes as large as $\Xi = 2.98$~PHz/strain for stretch or compression along the armchair direction, which is in the same order of magnitude that has been previously reported by some of us for the complex antisite defect (\vnnb) in hBN~\cite{Li2020}.
Such large deformation susceptibility emerges thanks to the two-dimensional nature of the setup and has the potential of leading us to the strong interaction between the vibrational modes and the color centers.

The general system Hamiltonian reads $H = H_{\rm S} +H_{\rm B} + H_{\rm I}$. We assume the color centers are optically driven at frequency $\omega_0$. Hence, in the frame that rotates with the laser drive frequency the Hamiltonian components are given by
\begin{subequations}
\begin{align}
H_{\rm S} &= \half\sum_k (\Delta_k\sigma^z_k +\Omega_k\sigma^x_k), \\
H_{\rm B} &= \sum_m \omega_m b_m^\dag b_m, \\ 
H_{\rm I} &= \half\sum_{k,m}\lambda_{k,m}\sigma^z_k(b_m +b_m^\dag),
\end{align}
\label{hamil}
\end{subequations}
where $\sigma^z = \trans{e}{e}-\trans{g}{g}$ and $\sigma^x = \trans{e}{g} +\trans{g}{e}$ are the Pauli matrices defined over the electronic ground and excited levels $\{\ket{g},\ket{e}\}$.
Here, $\Delta_k$ is detuning of the laser from the $k$th color center and $\Omega_k$ is the corresponding Rabi frequency.
$b_m$ ($b_m^\dag$) describe the annihilation (creation) of phonons in the $m$th flexural mode with the frequency $\omega_m$.
And $\lambda_{k,m}$ indicates the rate of coupling between the $k$th color center and $m$th mechanical mode.
The vibrational modes as well as the color centers are subject to decoherence. The vibrations are damped at the rate $\gamma_m =\omega_m/Q$ where $Q$ is the quality factor of flexural vibrations. In this work we shall assume a quality factor of $Q =5\times 10^5$, which is within reach for hBN resonators~\cite{CartamilBueno2017}.
As the main decoherence effect in the electronic states of the color centers we consider a relaxation that undergoes with the rate $\kappa_k$. The pure dephasing in the electronic structure is negligible thanks to the recent advanced experiments that control spectral diffusion of the emitters~\cite{Tran2019, White2021}, and thus, are omitted in this work.

\subsection{The setup}\label{sec:setup}
The monolayer structures of hBN exhibit excellent mechanical properties that guarantee high quality mechanical modes with large coupling strength to the electronic levels of the embedded color centers.
For a monolayer membrane of hBN the modulus of elasticity is $850$~GPa~\cite{Falin2017}.
The local strain experienced by a color center depends on the mode profile $\psi_m$ and its location on the BNNR.
The mode profiles for a strip with clamped-clamped boundary conditions are easily found by employing the elasticity theory of membranes, see Appendix~\ref{sec:elasticity}.
The coupling strength of a color center positioned at $z=z_k$ to the $m$th flexural normal mode is determined by the local strain imposed on it times the deformation susceptibility. For small deflections one finds
\begin{equation}
\lambda_{k,m} \approx \frac{\Xi}{2L^2}x_{{\rm zp},m}^2[\psi'_m(z_k)]^2,
\end{equation}
where $x_{\mathrm{zp},m} = \sqrt{\hbar/2M_m\omega_m}$ is the zero-point amplitude of the $m$th normal mode with effective mass $M_m$.

By considering a three-layer BNNR of length $L=1~\mu$m and width $w=3$~nm the coupling strength can reach values as high as $\lambda_{k,m}/2\pi = 7.2$~MHz.
Nonetheless, as it will shortly become clear when the bath degrees of freedom are eliminated from the dynamics effective coupling of the color centers to each other is determined by $G_{j,k} \approx \sum_m \lambda_{j,m}\lambda_{k,m}/\omega_m$, where we have neglected a damping contribution by assuming a high quality factor mechanical resonator.
Hence, the ratio of coupling strength and mode frequency $\lambda_{k,m}/\omega_m$ is the pivotal parameter is our setup.
The calculated values suggest that the value for modes with $m\geq 75$ is a thousand times smaller than the fundamental ratio which assumes the highest value among the spectrum: $\max\{\lambda_{k,1}\}/\omega_1 \approx 0.354$. Therefore, we truncate the sum at $m=75$ in our analyses by committing less than $1\%$ relative error.

In the rest of paper, we study the steady-state entangling scenario that can be envisaged in our system and study its immunity to the thermal noise of the vibrational modes.

%
%
\section{Reduced master equation}\label{sec:reduced}
To study the stationary entanglement scheme we first find an effective master equation that describes dynamics of the color centers by eliminating the vibrational degrees of freedom. This is performed by pursuing the standard projection formalism~\cite{Breuer2007}.
This way, the realistic case where the decoherence effects are present in the master equation $\dot\rho = \liov[\rho]$ is taken into account.
The Liouvillian is divided into three terms $\liov=\liov_{\rm S} +\liov_{\rm B} +\liov_{\rm I}$, where the components are given by
\begin{align*}
\liov_{\rm S}[\rho] &= -i[H_{\rm S},\rho] +\half\sum_k\kappa_k\mathcal{D}_{\sigma_k^-}[\rho], \\
\liov_{\rm B}[\rho] &= -i[H_{\rm B},\rho] +\half\sum_m \gamma_m\{\bar{n}_m \mathcal{D}_{b^\dag_m}[\rho] +(\bar{n}_m +1)\mathcal{D}_{b_m}[\rho] \}, \\
\liov_{\rm I}[\rho] &= -\tfrac{i}{2}\sum_{k,m}\lambda_{k,m}[\sigma_k^z(b_m +b_m^\dag ),\rho],
\end{align*}
where the dissipators are $\mathcal{D}_o[\rho] = 2o\rho o^\dag -o^\dag o \rho -\rho o^\dag o$.

We assume that the mechanical modes are in thermal equilibrium and the device temperature and their state is only slightly altered by their interaction with the color centers which is justified when $\lambda_{k,m} \ll \omega_m$. In our following numerical analysis we ensure the inequality always holds. Hence, we write $\rho(t) = r_{\rm ss}\otimes \mu(t)$, where $r_{\rm ss}$ is the bath thermal state and $\mu(t) = \mathrm{Tr}_{\rm B}\{\rho(t)\}$ is the reduced density matrix of color centers.
We define the projection $\proj\rho = r_{\rm ss}\otimes \mathrm{Tr}_{\rm B}\{\rho\}$ and its orthogonal complement $\comp \equiv \one -\proj$. The Nakajima-Zwanzig equation reads
\begin{equation}
\proj\dot\rho = \proj\liov\proj\rho +\proj\liov\int_0^t dt' e^{\comp\liov t'}\comp\liov\proj\rho(t-t').
\end{equation}
The evolution speed separation that is inherited in the three parts of the Liouvillian allows us to significantly simplify the above equation. Since we are assuming that the bath remains in its steady-state $\liov_{\rm B}\proj = 0$ and for trace preservation $\proj\liov_{\rm B}=0$. Therefore, we have $\proj\liov_{\rm B}\proj = \comp\liov_{\rm B}\proj = \proj\liov_{\rm B}\comp = 0$.
Furthermore, $\proj\liov_{\rm S}=\liov_{\rm S}\proj$ as they operate on different subspaces. As a result we have $\proj\liov_{\rm S}\comp = \comp \liov_{\rm S}\proj =0$.
The fact that the interaction Hamiltonian has no diagonal element in the eigenbasis of $H_{\rm B}$ gives $\proj\liov_{\rm I}\proj=0$.
Hence, the reduced master equation in Born approximation is
\begin{equation}
\dot\mu(t) = \liov_{\rm S}\mu(t) +\mathrm{Tr}_{\rm B}\Big\{\liov_{\rm I}\int_0^t\hspace{-1mm}dt'~e^{\liov_{\rm B}t'}\liov_{\rm I}[r_{\rm ss}\otimes\mu(t-t')] \Big\}.
\end{equation}
The second term in the above equation is simplified by employing Markov approximation which is equivalent to replacing $\mu(t-t')$ with $\mu(t)$.
This allows us to compute the bath correlation functions and perform the time integrations by setting $t\to\infty$, since we are interested in the steady-state entanglement.
The bath two-time correlation functions $C_m^+ \equiv \int_0^\infty\hspace{-1mm}dt~\mathrm{Tr}_{\rm B}\big\{x_m^\dag e^{\liov_{\rm B}t}r_{\rm ss}x_{m'} \big\}$ and $C_m^- \equiv \int_0^\infty\hspace{-1mm}dt~\mathrm{Tr}_{\rm B}\big\{x_me^{\liov_{\rm B}t}x_{m'} r_{\rm ss} \big\}$ with $x_m \equiv b_m +b_m^\dag$ are found by the quantum regression theorem~\cite{Carmichael1999}
\begin{equation}
C_m^\pm = \frac{1}{\frac{1}{4}\gamma_m^2 +\omega_m^2}\big[\gamma_m(\bar{n}_m +\half) \pm i\omega_m\big]\delta_{mm'}.
\end{equation}

Therefore, the reduced master equation becomes
\begin{align}
\dot\mu = \liov_{\rm S}[\mu] +\tfrac{1}{4}\sum_{j,k}\sum_m &\lambda_{j,m}\lambda_{k,m} \Big\{C_m^+(\sigma_j^z\mu\sigma_k^z -\mu\sigma_j^z\sigma_k^z) \nonumber\\
&+ C_m^-(\sigma_j^z\mu\sigma_k^z -\sigma_j^z\sigma_k^z\mu) \Big\}.
\end{align}
By rearranging the terms one finds the effective Hamiltonian
\begin{equation}
H_{\rm eff} = H_{\rm S} -\tfrac{1}{4}\sum_{j,k}G_{j,k}\sigma_j^z\sigma_k^z,
\label{eff}
\end{equation}
which explicitly shows bath-induced interaction between the color centers with the coupling rates $G_{j,k} \equiv \sum_m\lambda_{j,m}\lambda_{k,m}\Im\{C_m^+\}$.
Note that one could find the effective Hamiltonian by eliminating the interactions via a polaron transform~\cite{Contreras2008}.
This is accompanied by a dissipative coupling giving the following reduced master equation
\begin{align}
\label{reduced}
\dot\mu = -i[H_{\rm eff},\mu] &+\half\sum_k\kappa_k\diss_{\sigma_k^-}[\mu] \\\nonumber
&+\half\sum_{j,k}\Gamma_{j,k}\big(2\sigma_j^z\mu\sigma_k^z -\sigma_j^z\sigma_k^z\mu -\mu\sigma_j^z\sigma_k^z \big),
\end{align}
where $\Gamma_{j,k} \equiv \half\sum_m \lambda_{j,m}\lambda_{k,m}\Re\{C_m^+\}$ is the rate of dissipative coupling.
In the following, this master equation is employed for performing the numerical study via QuTiP package~\cite{Johansson2013}.

%
%
\section{Two-qubit entanglement}
To investigate the possibility of creating steady-state entanglement in our scheme we first perform a numerical simulation for the simplest case where only two color centers are present.
We consider the case that the color centers are positioned at $x_1=L/3$ and $x_2=2L/3$ of the BNNR. We shall later come back to the effect of error in the positioning of centers. This configuration gives an effective coupling of $G_{1,2}/2\pi \approx 2.2$~MHz.
\begin{figure}[tb]
\includegraphics[width=\columnwidth]{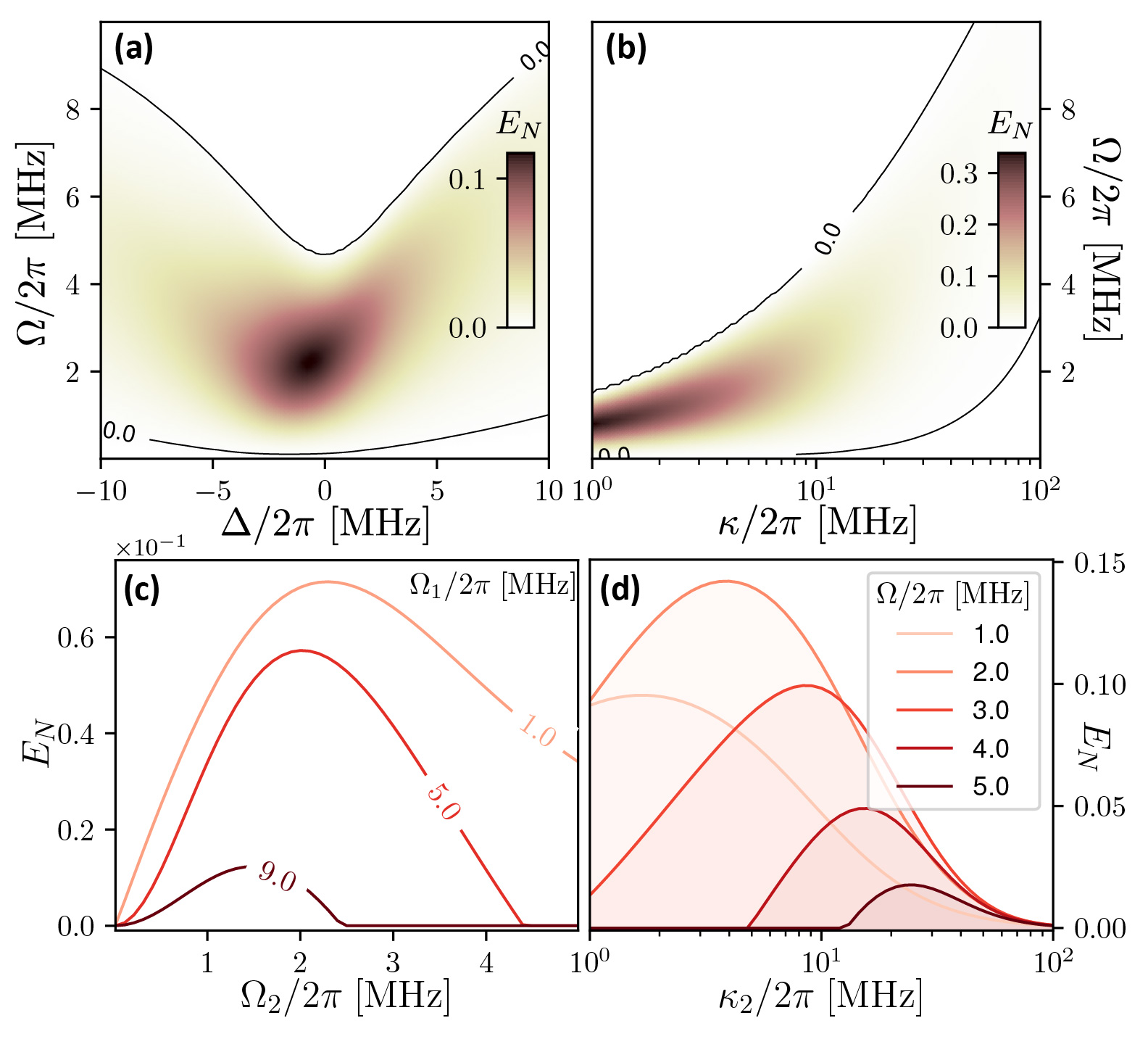}
\caption{%
Logarithmic negativity as a measure of entanglement between electronic levels of two color centers on a freestanding BNNR positioned at $L/3$ and $2L/3$:
(a) The amount of $E_N$ at different values of laser drive detuning and Rabi frequencies when the decay rate is fixed at $\kappa/2\pi =10$~MHz.
(b) Variations of $E_N$ versus the color center decay rate when they are driven on resonance at different Rabi frequencies.
(c) Effect of dipole moment orientation for two defects of the same kind: $E_N$ versus Rabi frequency of qubit-2 for three different $\Omega_1$ values.
(d) Robustness of the steady-state entanglement for two qubits with different decay rates. Qubits with decay rates within various ranges can be entangled by choosing proper Rabi frequencies. Here, decay rate of qubit-1 is fixed at $\kappa_1/2\pi=10$~MHz and we take $\Omega\equiv\Omega_1=\Omega_2$.
The mechanical quality factor is set to $Q=5\times 10^5$.
}%
\label{fig:two}%
\end{figure}
The color centers are driven by a single laser as the length of BNNR that we are considering here stretches to the diffraction limit.
Note that even though the coherent couplings do not depend on $L$ for a doubly-clamped configuration, $\Gamma_{j,k}$ increases with the length of BNNR. Hence, we have chosen a minimum length that is yet experimentally feasible.
Assuming that the color centers are of the same type, e.g. \vb\ one can assume equal detunings $\Delta_1=\Delta_2\equiv\Delta$, decay rates $\kappa_1=\kappa_2\equiv\kappa$, and Rabi frequencies $\Omega_1=\Omega_2\equiv \Omega$.
On the other hand, the \textit{ab initio} calculations of the current work on \vb\ as well as a similar analysis performed for \vnnb\ suggest that the strain susceptibly of both defect kinds is about the same magnitude. Nonetheless, the transition frequency and decay rate of these defects are very different.
While the mismatch in resonance frequencies can be compensated for by employing a two-tone laser drive, the latter is intrinsic. Therefore, one also needs to consider the case of different emitter decay rates $\kappa_1\neq\kappa_2$.
Furthermore, in the case of \vnnb\ centers the two qubits may assume different dipole polarization orientations, and thus, they feel different electric fields from the driving laser. Hence, the Rabi frequencies can also be different.

Fig.~\ref{fig:two} shows the results for exactly positioned defects.
First, the optimal detuning and Rabi frequencies are determined through the values that maximize entanglement of the two qubits with the same drive parameters. In Fig.~\ref{fig:two}(a) the density plot of logarithmic negativity $E_N$ is presented, signifying the importance of working parameters [see Appendix~\ref{sec:measures} for the definition of entanglement measures used in this work]. Here, we have fixed the decay rate at $\kappa/2\pi = 10$~MHz and assume an ambient temperature of $T=30$~mK. The maximum attainable steady state entanglement in this case is $\max\{E_N\}\approx 0.1$.
To see how different variants of color centers of the same kind are entangled to each other in this configuration, we plot $E_N$ as a function of Rabi frequency and decay rate. One observes that the lower decay rates, result in higher steady-state entanglements, provided operating the laser at the optimal $\Omega$ value.
\begin{figure}[b]
\includegraphics[width=\columnwidth]{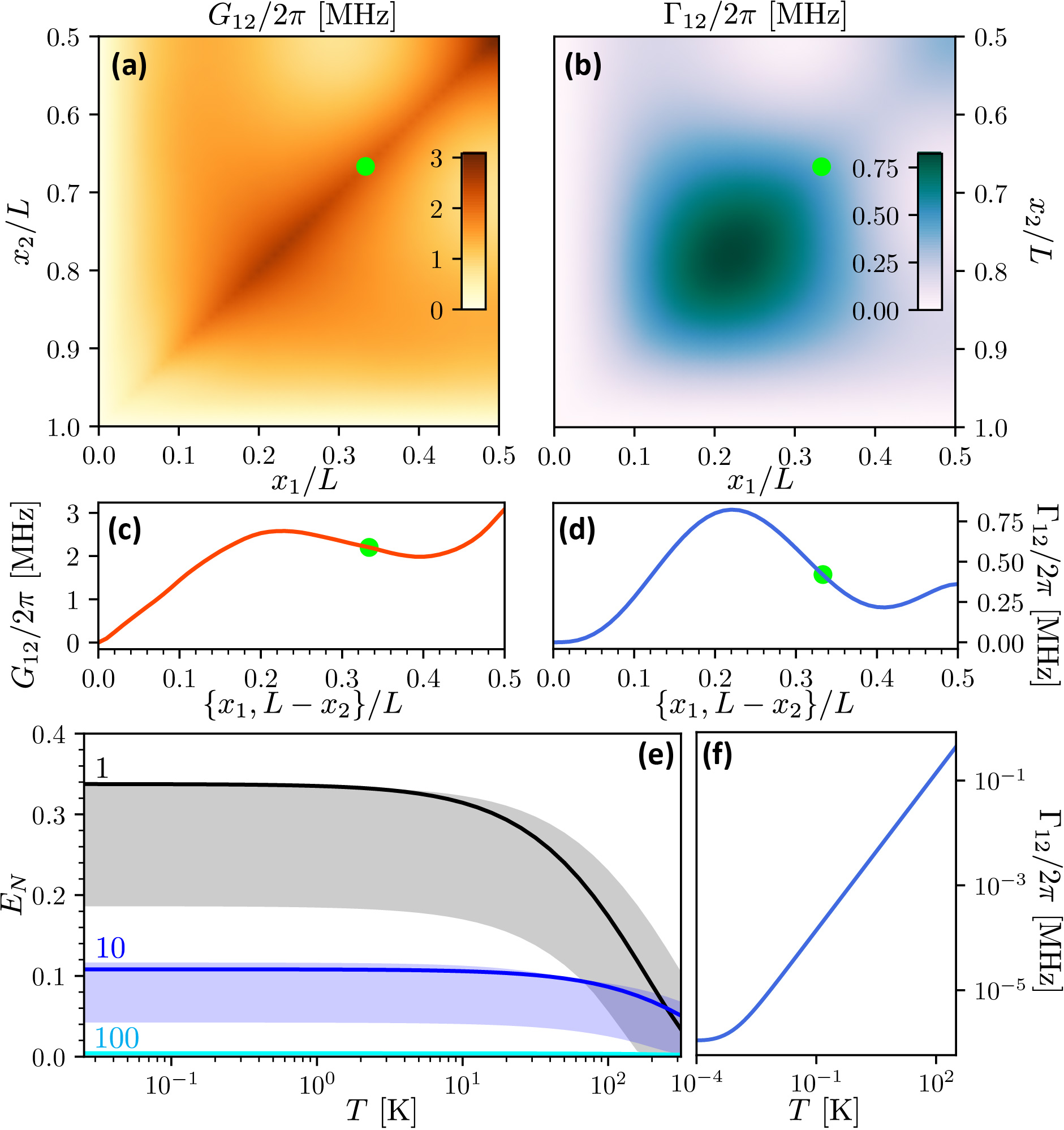}
\caption{%
Dependence of (a) $G_{1,2}$ the coherent and (b) $\Gamma_{1,2}$ the dissipative coupling rates between two color centers on their position along a BNNR of $L=1~\mu$m. The latter is calculated at room temperature. In (c) and (d) the diagonal values, i.e. the quantities at $x_1=L-x_2$ are shown. The green dots pinpoint the positions and parameters studied in this work.
(e) Effect of thermal noise and mis-positioning on the entanglement for two color centers. The bold lines show the entanglement as a function of temperature when exactly positioned at the one-third points (the green dots). The shades give the value when a ten percent error in the positions are taken into account, see the text for further details. The numbers next to each line show the decay rate of centers, $\kappa/2\pi$~[MHz]. 
(f) Variations of the dissipative coupling rate with the ambient temperature for $\{x_1,x_2\}=\{\frac{L}{3},\frac{2L}{3}\}$.
}%
\label{fig:noise}%
\end{figure}

Next, we investigate the effect of different Rabi frequencies with a single optical drive. The difference is basically due to various orientations of the defects electric dipole moment $\mathbf{p}$ with respect to optical polarization $\bm{\varepsilon}$ such that $\Omega_k \propto \mathbf{p}_k\cdot\bm{\varepsilon}$.
This, in turn, can stem from the different defect kinds or different dipole orientation in color centers of the same kind, see Fig.~\ref{fig:scheme} for the geometry.
Fig.~\ref{fig:two}(c) shows the logarithmic negativity of two-qubit steady-state against their Rabi frequencies when driven on resonance ($\Delta=0$). Note that $\Omega_k$ can also take negative values. Nevertheless, because of the symmetry in the sign dependence, it is enough to look at the positive values.
We notice that even for different polarization orientations one can engineer the optimal optical polarization such that the two color centers get stationary entanglement with tolerance on deviations $|\Omega_1-\Omega_2|$ which covers a quite wide parameter region.
We now consider the case of two different kinds of point defects that have similar strain susceptibility but different decay rates. Note that the on-resonance drive can be attained by two-frequency laser drive and the optimal Rabi frequencies are obtained by engineering the laser polarizations. But the decay rate is intrinsic and only slightly manipulable, e.g. by Purcell effect.
Hence, in Fig.~\ref{fig:two}(d) we show the results for entanglement of emitters with different decay rates and find that the scheme is rather robust regarding the color center varieties, provided they both exhibit appreciable deformation potentials.
Moreover, the choice of $\Omega$ is crucial for covering the desired $\kappa$ ranges.

As we mentioned before the position of color centers determine the strength of their mutual interaction. To illustrate this the density plots of $G_{1,2}$ as well as $\Gamma_{1,2}$ are shown as a function of the position of the centers in Fig.~\ref{fig:coupling}(a) and (b), respectively. The coupling strength depends on the mode profiles that determine the local strain exerted on the defect and their frequency.
We notice that the coherent coupling $G_{1,2}$ is only appreciable when the qubits are equally distanced from their respective ends [Fig.~\ref{fig:noise}(a)]. That is, along the line with $x_1=L-x_2$. The variations of $G_{1,2}$ on this line is separately shown in Fig.~\ref{fig:noise}(c). And the coupling rate is maximized when both color centers are close to the middle point of the BNNR.
The dissipative coupling rate behaves rather differently and assumes its highest value about a wider range of qubit positions centered around the first and third quarters of the BNNR [Fig.~\ref{fig:noise}(b) and (d)].
As it shortly becomes clear, the dissipative coupling mostly has a destructive effect on the steady-state entanglement. Hence, one exploits these different distributions for enhancing the entanglement.
Meanwhile, any error in positioning of the color centers can significantly affect the way they are coupled; it may result in a smaller $G_{1,2}$ accompanied with a higher $\Gamma_{1,2}$, which inevitably reduces the amount of $E_N$.
Effect of such mis-positionings and the thermal noise---as two main imperfections---on the entanglement is summarized in Fig.~\ref{fig:noise}(e) where we plot $E_N$ as a function of temperature. In this plot, the value of entanglement is shown for a perfect positioning at $\{x_1,x_2\}=\{L/3,2L/3\}$ (bold lines). Alongside, a shaded area is shown that gives variations in the logarithmic negativity when $10\%$ error ($\Delta x_k = 100$~nm) is taken into account in the calculations.
To generate the shaded areas in the plot, we have taken a hundred random samples of positions within the range $\{x_1,x_2\} = \{\frac{L}{3}\pm\frac{L}{10},\frac{2L}{3}\pm\frac{L}{10}\}$ for every value of temperature.
The three different line colors in Fig.~\ref{fig:noise}(e) correspond to different decay rate values $\kappa$. In both of them the color centers are driven on resonance $\Delta=0$ with an optimal Rabi frequency.
One observes that the entanglement is robust against the ambient temperature and survives even at room temperature, provided the relaxation rate of the qubits is large enough.
The main source for the loss of entanglement is the enhanced decoherence rates induced by the thermal vibrational bath, $\Gamma_{\!j,k}$, with temperature. The curve in Fig.~\ref{fig:noise}(f) shows that these decoherence rates rapidly increase with the rise of temperature and even become comparable to the coherent coupling strength $G_{1,2}$ at room temperature.

%
%
\section{Graph states}
Graph states can serve as a resource both in one-way quantum computation protocols and quantum metrology. Universality of the computations in a cluster state are guaranteed when a square lattice is formed in which each qubit experiences a $\sigma^z_j\sigma^z_k$ interaction to its four nearest neighbors~\cite{Briegel2001}.
Graph states with a higher degree of complexity are beneficial in quantum metrology where a full graph with high enough entanglement can break the standard quantum limit and reach the Heisenberg limit~\cite{Giovannetti2011}.
The effective Hamiltonian in Eq.~\eqref{eff} suggests that the basic ingredients for both purposes are present in our proposed scheme. To assess the possibility of creating steady-state multipartite entanglement among several color centers, we perform a numerical evaluation on the reduced master equation~\eqref{reduced}.
By setting left hand side of~\eqref{reduced} equal to zero, one find the stationary state. We investigate its quantum properties by evaluating two parameters: the genuine multipartite entanglement~\cite{Coffman2000, Ou2007} as well as quantum Fisher information (QFI)~\cite{Braunstein1994}.
The definition of these measures is provided in Appendix~\ref{sec:measures}.
In the following study, we consider three and more number of color centers and assume that the defects are positioned in a way that the coupling strength of them to each other is maximized. We numerically find these optimal positions of $N$ point defects $\{x_k\}_N$ by assigning a `district' to each color center such that $\frac{k-1}{N}+\xi<x_k/L<\frac{k}{N}-\xi$, where $\xi$ is a small fraction of district length to ensure that no two qubits occupy the same position. Here, we set $\xi =0.1$.
\begin{figure}[tb]
\includegraphics[width=\columnwidth]{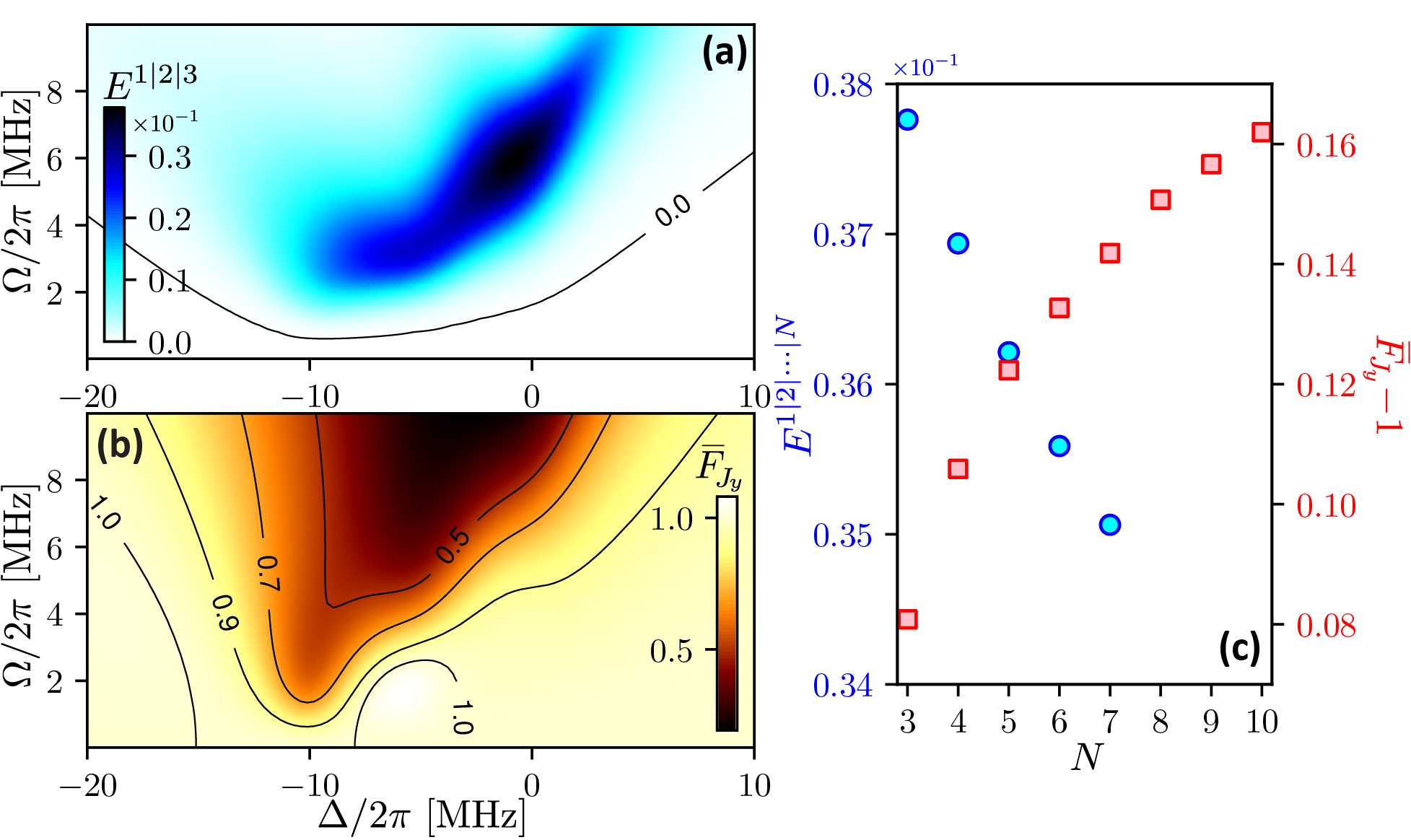}
\caption{%
Entanglement properties of a multi-particle system.
(a) The genuine multipartite entanglements as a function of laser detuning and Rabi frequency.
(b) The mean quantum Fisher information where $\bar{F}_{J_y}>1$ signals crossing the standard quantum limit.
(c) Variations of the optimal multipartite entanglement $E^{1|2|\cdots|N}$ and boost in the Fisher information $\bar{F}_{\!J_y}-1$ with the number of color centers.
In these plots the parameters are $T=30$~mK, $\kappa/2\pi = 10$~MHz, and the optimal position for the color centers is considered such that their mutual coupling rates are maximized, see the text for further details.
}
\label{fig:cluster}
\end{figure}

In Fig.~\ref{fig:cluster} we show the multipartite features of a group of color centers. A genuine multipartite entanglement at the steady state of three-qubit and more systems is attainable with optimal choice of optical drive parameters $\Delta$ and $\Omega$.
The numerical analyses in this section are performed by considering $\kappa/2\pi = 10$~MHz. For the ease of comparison, it is worth mentioning that for an ideal Greenberger-Norne-Zeilinger (GHZ) state which is given by $\ket{\rm GHZ}_N = (\ket{e}^{\otimes N} +\ket{g}^{\otimes N})/\sqrt{2}$ the entanglement measure employed here gives $E^{1|2|\cdots|N}(\ket{\rm GHZ}_N)=1$. Fig.~\ref{fig:cluster}(a) presents $E^{1|2|3}$ as a function of laser detuning and Rabi frequency when the color centers are located at $\{x_k\}_3 = \{0.30, 0.36, 0.69\}\times L$. The optimal location of higher number of color centers are numerically found, e.g. $\{x_k\}_4 = \{0.22, 0.27, 0.72, 0.77\}\times L$ and $\{x_k\}_5 = \{0.16, 0.22, 0.42, 0.76, 0.82\}\times L$, and so on.

As another measure of entanglement for assessing a graph state we use mean quantum Fisher information which is bounded from below and above through $0\leq \bar{F}_O \leq N$ for an $N$-partite system. The upper bound is the Heisenberg limit, yet for a separable coherent state one expects $\bar{F}_O=1$ when a proper measurement operator $O$ is considered. Here, we only consider linear measure operators of $O = J_n \equiv n_x J_x +n_y J_y +n_z J_z$ where $J_\alpha \equiv \half\sum_{i=1}^N \sigma_i^\alpha$ with $(\alpha =x,y,z)$ are the collective spin operators and $\hat{n} = (n_x,n_y,n_z)$ is a unit vector.
For a given state one determines $\hat{n}$ such that the value of $\bar{F}_{J_n}$ gets maximized.
 For example, for the state $\ket{\rm GHZ}_N$ we find $\bar{F}_{J_z} = N$, which touches the Heisenberg limit.
 In Fig.~\ref{fig:cluster}(b) we present QFI of the tripartite system in terms of $O = J_y$. Given the steady-state property and high decay rate of the qubits $\bar{F}_{J_y}$ is even below the standard quantum limit (SQL) in most parameter regions.
 Nonetheless, one still is able to identify areas that $\bar{F}_{J_y}>1$, though slightly, which signals passing the SQL as it is expected from a graph state~\cite{Shettell2020} as well as a multiparticle entangled state~\cite{Hyllus2012}.
 This behavior is studied for larger systems and the optimal values are shown in Fig.~\ref{fig:cluster}(c) where the genuine multipartite entanglement $E^{1|2|\cdots|N}$ as well as the boost in quantum Fisher information $\bar{F}_{\!J_y}-1$ are given for three- to ten-particle systems. Note that because of the expensive computations we only calculate the multipartite entanglement for up to $7$-particle systems.
Surprisingly, even though the maximum value of $E^{1|2|\cdots|N}$ decreases with $N$, the enhancement in the QFI exhibits a growing behavior. This result suggests that a BNNR with almost evenly positioned color centers has the potential for quantum sensing, e.g. magnetic fields.

It is worth mentioning that the rather small multipartite entanglement and boost in the Fisher information that are found in our setup stem from the realistic noise effects that are considered in our analyses. However, note that it is also partly because of the steady-state nature of these features, see Refs.~\cite{Huelga2007, Valle2011}.

%
%
%
\section{Phonon superradiance}
In this section, we theoretically study the driven-dissipative phase transition in the scheme proposed above, where the color centers and one of the BNNR vibrational modes experience a normal-to-superradiant transition.

The superradiance phase transition of a bosonic field in interaction with an assembly of two-level atoms was first studied by Dicke in 1954~\cite{Dicke}.
The bosons in Dicke model can consist of photon or phonon modes. The superradiance is a collective effect that describes the emission of coherent photons or phonons by a large number of atoms when interaction strength is greater than a threshold coupling, the critical coupling.
In the past few years, the photon superradiance transition in the Dicke model has been largely investigated both theoretically and experimentally such as by using stimulated Raman emission to couple the atoms with cavity photons~\cite{Dimer2007, Zhiqiang:17, Zhang2}, driving a gas of thermal atoms trapped inside a cavity by an external coherent pump~\cite{Domokos2002,Black2003}, employing superconducting qubits in interaction with a common microwave resonator~\cite{Jaako, Bamba2, Viehmann, nataf2010no}, as well as with quantum dots as artificial atoms~\cite{scheibner}, and nitrogen-vacancy centers in diamond~\cite{Zou2014}.
 Moreover, there have been several proposals for realizing superradiance transition by employing motional degrees of freedom of a mechanical vibrational such as coupling the electronic states of trapped ions to their center of mass motion~\cite{Genway, Safavi, Wang}, and using Bose-Einstein condensates where the spin-orbit coupling creates a collective interaction between internal spin state and motional degrees of freedom~\cite{baumann2010dicke, hamner2014dicke}.
 The mentioned observations have signified the interest and motivation for further investigating the collective effects in the presence of a phonon mode. The phonon superradiance have been observed in various systems as nanomagnets \cite{Chudnovsky} and quantum dots~\cite{Brandes}.

To inquire the phonon superradiance in the BNNR, we rearrange the Hamiltonians in Eqs.~\eqref{hamil} and include a `selected' vibrational mode and its coupling to the qubits in the system Hamiltonian. After following similar procedure explained in Sec.~\ref{sec:reduced} for elimination of the rest of the vibrational spectrum one arrives at the following effective Hamiltonian which is also known as Dicke-Ising model
\begin{align}
\label{HDIC}
H_{\rm D} =& \half\sum_k (\Delta_k\sigma^z_k +\Omega_k\sigma^x_k) -\tfrac{1}{4}\sum_{j,k}G_{j,k}\sigma_j^z\sigma_k^z \nonumber\\
& +\omega a^\dag a +\tfrac{1}{2}\sum_k \lambda_k\sigma^z_k (a+a^\dag), 
\end{align}
where $a^\dag$ ($a$) is the creation (annihilation) operator of the selected mode and $\omega$ is its angular frequency.
This is accompanied with a reduced master equation which basically the same as Eq.~\eqref{reduced} but with $H_{\rm D}$ replacing $H_{\rm eff}$ and the dissipation of the boson mode $\half \gamma\{\overline{N} \mathcal{D}_{a^\dag}[\mu] +(\overline{N} +1)\mathcal{D}_{a}[\mu] \}$ included in the Liouvillian, where $\overline{N}$ is the thermal occupation number of the boson.
One must also keep in mind that the contribution of the \textit{selected} mode is deducted from the effective coupling $G_{j,k}$ and decoherence $\Gamma_{j,k}$ rates.

We are interested in the steady-state value of $a$ as the order parameter that signals occurrence of the phase transition. Hence, from the reduced master equation one easily finds
\begin{equation}
\mean{a}_{\rm ss} =\frac{\sum_k\lambda_k \mean{\sigma^z_k}_{\rm ss}}{i\gamma-2\omega}.
\label{ass}
\end{equation}
Apparently, this value depends on the single-qubit steady-state expectation values. The dynamics of the color centers, in turn, is more complicated. 
We thus apply the mean-field approximation to eliminate the bosonic mode from the qubit dynamics so the coupling terms read
\begin{equation}
\sigma_k^z(a+a^\dag)\xrightarrow[]{\text{mean-field}}2\sigma_k^z\Re\{\mean{a}_{\rm ss}\}.
\end{equation}
Therefore, the resulting master equation becomes the same as Eq.~\eqref{reduced} but with modified detuning parameters $\Delta_k\to\tilde\Delta_k\equiv \Delta_k +2\lambda_k\Re\{\mean{a}\}$, where we have dropped the `ss' subscript for the convenience. We denote this modified reduced master equation by $\dot\mu=\liov_{\rm D}[\mu]$.
It must be emphasized that the phase transition disappears as soon as the drive term is removed from Hamiltonian \eqref{HDIC} since, in that case, no excitation exchange between the boson and qubits is possible.

\begin{figure}[tb]
\includegraphics[width=\columnwidth]{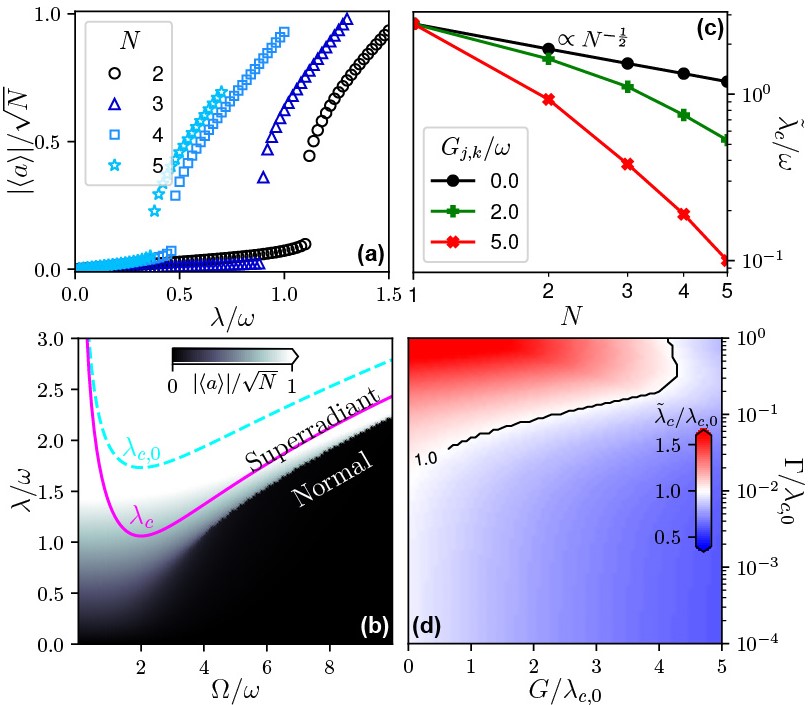}
\caption{%
The phonon superradiance:
(a) Variations of the vibrational mode amplitude $\mean{a}/\sqrt{N}$ as the order parameter with the qubit-boson coupling strength for a setup with $N=\{2,3,4,5\}$ color centers. The selected mode frequency in each case is $\omega=\omega_{N+1}=2\pi\times\{24.49, 43.42, 60.48, 84.48\}$~MHz. The inter-qubit coupling strengths are numerically determined in each case.
(b) Phase diagram for the two qubit case.
(c) Dependence of the critical point on the number of color centers and their mutual couplings.
(d) The deviation of critical point from the non-interacting qubits case in Eq.~\eqref{critical} with the coherent $G$ and incoherent $\Gamma$ coupling rates in a system with two color centers.
In (a) and (c) $\Omega=5\omega$, while in (d) $\Omega=9\omega$.
}%
\label{fig:dicke}
\end{figure}

In order to determine the critical point of the system, we first assume identical parameters for the color centers ($\Delta_k=\Delta=0$, $\Omega_k=\Omega$, $\lambda_k=\lambda$, $\kappa_k=\kappa$, $G_{j,k}=G$, and $\Gamma_{k,k}=\Gamma$) and neglect the incoherent coupling between the color centers $\Gamma_{j,k}=0 \text{ for } j\neq k$.
We derive the equations of motion for the mean-field dynamics from the reduced master equation and perform a linear stability analysis on them by employing Routh-Hurwitz stability criterion to find the critical coupling at zero temperature as~\cite{Gilmore1978, Gelhausen2017, Kirton2017}
\begin{equation}
\lambda_c = \Big[\frac{(\frac{1}{4}\gamma^2+\omega^2)((\frac{1}{2}\Gamma+2\kappa)^2+\Omega^2-\frac{1}{2}G\Omega)}{N\omega\Omega}\Big]^{\half},
\label{critical}
\end{equation}
which reduces to that of standard Dicke model $\lambda_{c,0} \equiv \sqrt{\omega(4\kappa^2+\Omega^2)/N\Omega}$ for $G=\Gamma=0$.
Note that the critical coupling tends to infinity as $\Omega\to0$, further supporting the fact that without a drive the system would not experience a phase transition.
Because of the approximations that are made in the path of arriving at the above analytical relation for the critical coupling and the absence of effects such as the mutual incoherent interaction among the color centers motivates us to use a numerical approach for determining the critical point. Therefore, in the following study we employ an iterative method with the semiclassical approach for tracking the phase transition in the steady-state of our proposed scheme.
In our method, a nontrivial initial guess is considered for the boson field $\mean{a}$. Then the steady-state of the qubit system is numerically found by solving the reduced master equation $\liov_{\rm D}[\mu_{\rm ss}]=0$, which is tractable thanks to the finite number of qubits we are considering in our study. The qubit steady-state expectation values are thus computed and plugged back in Eq.~\eqref{ass} to find a renewed value for $\mean{a}$. The process is repeated until the value of $\mean{a}$ converges within the tolerance.
By slowly increasing the value of coupling rate our method reveals a phase transition from the normal phase with $\abs{\mean{a}}=0$ to the superradiant phase with $\abs{\mean{a}}>0$ at a critical coupling which we denote by $\tilde\lambda_c$.
See Appendix~\ref{sec:method} for a discussion on the convergence in our method.

In our investigation, we consider $N$ color centers positioned on a BNNR such that they all are identically coupled to the selected vibrational mode $\lambda_k=\lambda$. This can be attained by adjusting the position of qubits at points with identical local strain, which in turn are identified from the mode profiles. We find that the $n$th normal mode supports $n-1$ local equal strain maxima. Hence, in our study the selected vibration mode is set as $\omega = \omega_{N+1}$ for an $N$-qubit system, since this guarantees the highest $\lambda/\omega$ ratio which is crucial for the phase transition.
To study the phonon superradiance, we compute the normalized steady-state expectation value, $\abs{\mean{a}}/\sqrt{N}$, for various system parameters with the method described above for a monolayer BNNR membrane with properties given in Sec.~\ref{sec:setup} at the temperature $T=10$~mK. Furthermore, a fixed decay rate of $\kappa/2\pi = 20$~MHz for the color centers is taken into account.

In Figs.~\ref{fig:dicke}(a) The variations of the order parameter $\mean{a}/\sqrt{N}$ with the coupling rate is shown. The system exhibits the occurrence of phonon superradiance in the available system parameters when two to five color centers are employed. Note that the plots are only drawn up to the highest available $\lambda$ in the corresponding system, e.g. $\lambda/\omega = 0.73$ for $N=5$. Moreover, the corresponding $G_{j,k}$ and $\Gamma_{j,k}$ values are taken into account in the numerical calculations.
From these plots one also notices that the transition is rather smooth due to the finite decay rates.
Fig.~\ref{fig:dicke}(b) presents a typical phase diagram in our scheme for the case of a system with only two color centers interacting with each other at $G_{j,k}/\omega=5$ and $\Gamma_{j,k}/\omega=6\times10^{-5}$ which are found from the geometry and setup properties. We observe a sharp and clear transition from normal to superradiant phase for Rabi frequencies much larger than the qubit decay rate $\Omega \gg \kappa$, where $\kappa\approx\omega$ in this case. However, the border of the two phases sets to fade out for Rabi frequencies comparable to the decay rate $\Omega \sim \kappa$.
The magenta (dashed cyan) line indicates the approximate analytical critical coupling $\lambda_c$ ($\lambda_{c,0}$) found in the presence (absence) of the inter-qubit interactions.

To study the effect of system size on the critical behavior of the system we compute the critical coupling $\tilde\lambda_c$ for different number of qubits $N$ at three different inter-qubit coupling strengths and present the log-log plots in Fig.~\ref{fig:dicke}(c). Except for the case of $G_{j,k}=0$, which one finds $\tilde\lambda_c\propto N^{-\frac{1}{2}}$ as predicted by Eq.~\eqref{critical}, the finite coupling between the color centers leads to the onset of superradiance at smaller coupling rates.
We, nonetheless, note that by increasing $G$ the inter-qubit decoherence $\Gamma$ is also increased. Hence, we present the competition between the coherent and incoherent coupling among two color centers ($N=2$) in Fig.~\ref{fig:dicke}(d) by computing the critical coupling normalized to the one predicted by a standard open Dicke model $\lambda_{c,0}$.
For small enough decoherence rates $\Gamma$ the coherent coupling shifts the superradiance to the smaller qubit-boson couplings, see the darker shades of blue at the lower right corner of the density plot.
Even though this behavior is moderated or even reversed by larger $\Gamma$ values when $G\lesssim 4\lambda_{c,0}$, a slightly larger value of $G$ can still make the superradiance available at more affordable values of $\lambda$.

It is worth noting that our results provide the proof of principles for the emergence of a phase transition in the driven-dissipative finite size Dicke-Ising model~\cite{Konya2012, Hwang2018} implementable by color centers on a hBN membrane.
Finally, our scheme circumvents the ``no-go theorem'', which states the superradiant transition cannot be obtained using only dipole couplings between two-level atoms and photon field mode~\cite{Rzacewski1991}, since here the color centers are coupled strongly to the bosonic field through the local strain.
Furthermore, in our driven-dissipative scheme $\lambda_c\propto\sqrt{\Omega}$, and thus, the phase transition is facilitated thanks to the laser drives. Adding to this the inter-qubit couplings it is possible to observe the transition at affordable coupling rates.

%
%
\section{Summary and conclusion}
In summary, we have computed the strain susceptibility of boron vacancy defects in monolayer hexagonal boron nitride membrane via \textit{ab initio} calculations. Our study shows that a significant sensitivity to the local strain exhibits in the electronic levels of these defects.
We have exploited this exceptional property to propose a device where the flexural modes of a freestanding BNNR serve as mediators in the coupling of two and several quantum emitters.
As an example for applications of such device, we have studied the long living entanglement of the color centers.
Our analyses suggest that one is able to create steady-state entanglement among the color centers which is robust against different imperfections, including the axial and azimuthal positioning of the centers, as well as the thermal noise imposed through the vibrations.

By computing the genuine multipartite entanglement and quantum Fisher information in the case of three- and multi-qubit systems we have numerically proved the principle of generating graph-states that are essentially beneficial for enhanced quantum sensing as well as one-way quantum computation.
Given the experimental feasibility of our scheme this work paves the way towards long-lived electronic entangling of solid state emitters.

Moreover, we have investigated implementation of the Dicke-Ising model based on the hBN color centers and we have proved the feasibility of steady-state phonon superradiance observation in our scheme. The phase transition is available thanks to the high qubit-boson coupling as well as the coupling of qubits to each other. These both stem from the high strain susceptibility of \vb\ and \vnnb\ point defects and good mechanical properties of hBN membranes. 


%
%
\begin{acknowledgments}
FTT and MA acknowledge the supported by Iran Science Elites Federation.
\end{acknowledgments}

\appendix
\begin{figure*}[tb]
\includegraphics[width=1.0\columnwidth]{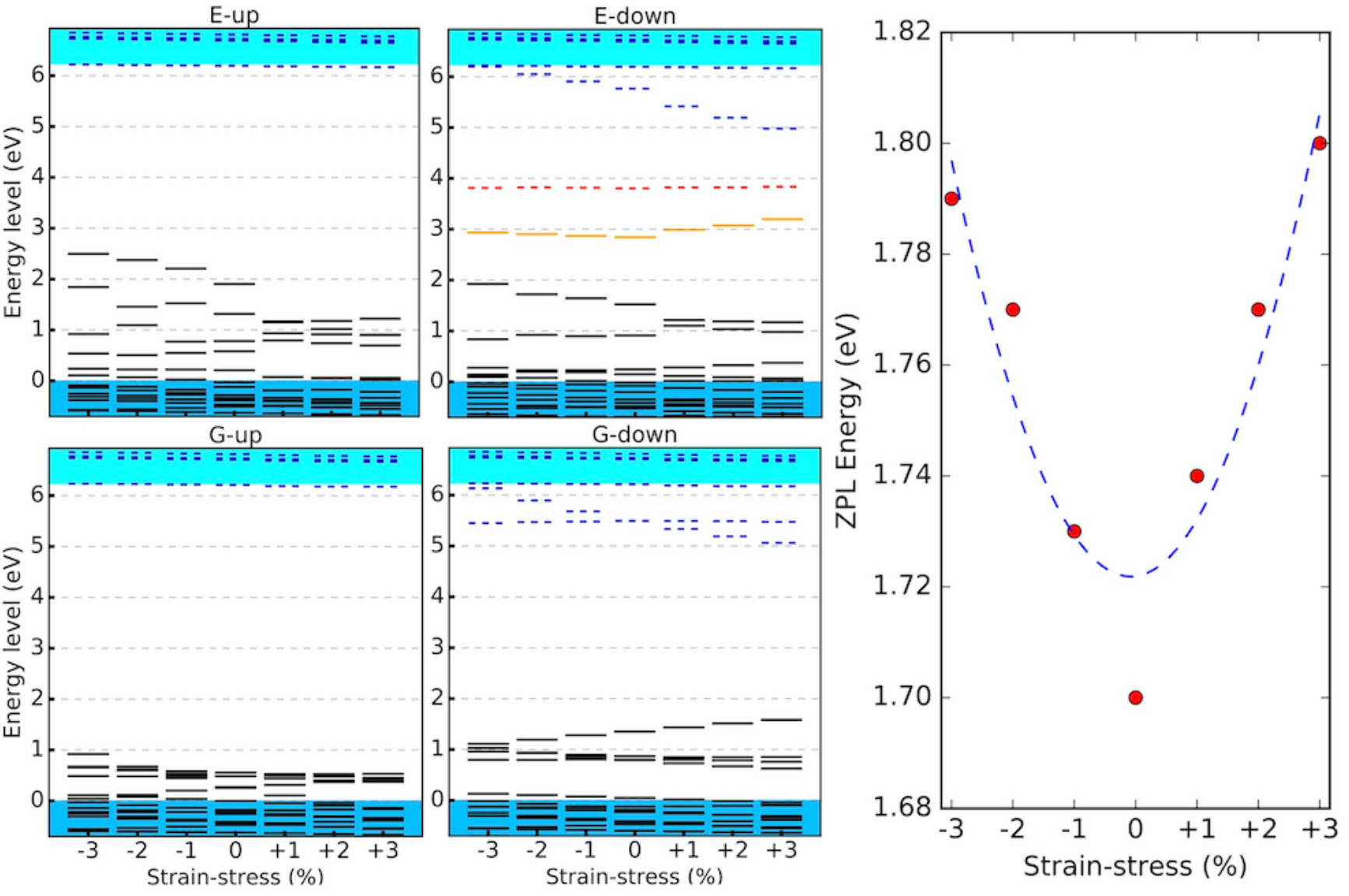}
\includegraphics[width=1.0\columnwidth]{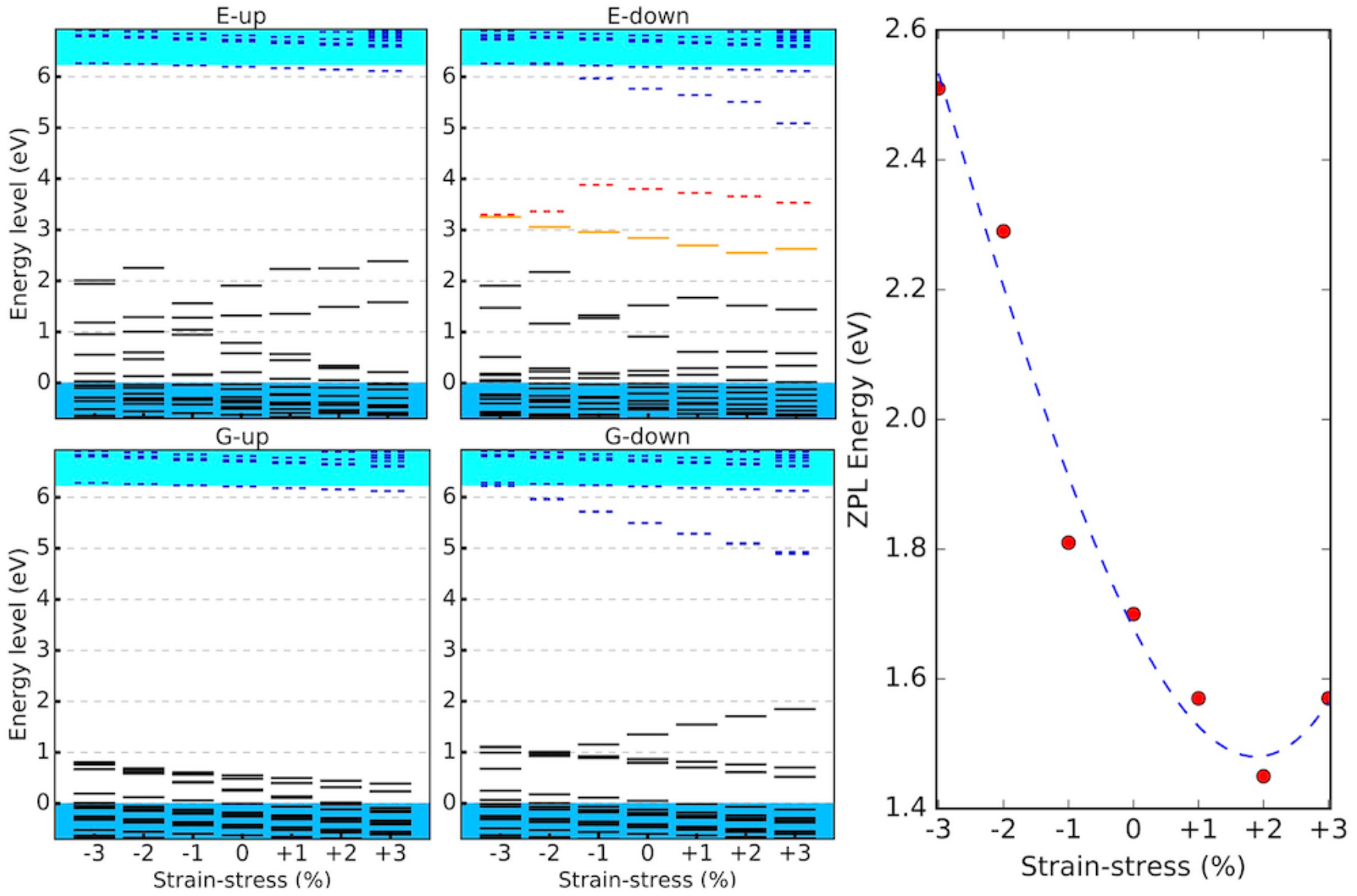}
\caption{%
The same plot as Fig.~\ref{fig:scheme}(b) and (c) for uniaxial strain along zigzag direction (left) and for equal-biaxial strain (right).
}
\label{fig:zigzag}
\end{figure*}
%

\section{Complementary electronic structure study}\label{sec:DFTstudy}
In this appendix, we provide our DFT calculations on the strain applied in the zigzag direction as well as a biaxial strain.
In the case of the zigzag direction we encounter a parabolic behavior for the ZPL energy, see Fig.~\ref{fig:zigzag}. In this direction range of the changes are small (almost 100 meV) and zero strain posses smallest ZPL energy.
In the biaxial stress-strain which is a combination of both uniaxial strains (armchair and zigzag), the trend of the ZPL energy is expected to manifest a mixture of zigzag and armchair uniaxial strains as it is shown in Fig.~\ref{fig:zigzag}.

In the \vb\ defect the $^{3}\!\textrm{A}'_{2}$ ground state with $D_{3h}$ point group symmetry suffers from Jahn-Teller distortion and  after a change in the geometry, it will find lower energy. Due to this fact, we think that $^{3}\!\textrm{E}'\rightarrow\ ^{3}\!\textrm{A}'_{2}$ transition reaches higher ZPL energy in the experiment. However, we do not anticipate the systems under strain to experience Jahn-Teller distortion because strain breaks degenerate states.
In a recent report, applying 3.7\%  strain to an SPE along zigzag direction shifted the ZPL energy up to 49 meV whereas in another SPE 5.5\% strain along the armchair direction shifted the ZPL about 65 meV~\cite{Mendelson2020}.
In a system without strain in the ground state, due to the existence of degeneracies in the levels, the Jahn-Teller  effect occurs. With the occurrence of the Jahn-Teller effect, the energy of the ground state is reduced due to the breaking of the degeneracy, so the ZPL energy increases. By taking into account the Jahn-Teller distortion, we believe that among these deformations the applied strain along the zigzag direction resembles experimental results. 

\section{Elasticity of BNNR}\label{sec:elasticity}
In this appendix, we provide details on the elasticity equations that have been employed in our work to describe the vibrational properties of the BNNRs.
The flexural dynamics of a membrane are thoroughly studied in Ref.~\cite{Landau1975}. The dynamics of such configuration for transverse displacements $\xi(z,t)$ that are much smaller than the length $L$  is described by~\cite{Wang2006}
\begin{equation}
\rho_{\rm B}h\partial_t^2\xi = -D\partial_z^4\xi +T\partial_z^2\xi,
\label{euler}
\end{equation}
where $E$ is bulk Young's modulus, $T=T_0+\Delta T$ is the built-in tension, and $D=Eh^3/12(1-\sigma^2)$ with $h$ the membrane thickness and $\sigma$ the Poisson ratio. Here, $\rho_{\rm B}$ is the bulk mass density.
By inserting the ansatz $\xi(z,t)=\psi(z)e^{-i\omega t}$ in Eq.~\eqref{euler} the eigenvalue equation is found that by solving it one finds the normal mode profiles as well as their corresponding frequencies
\begin{equation}
\rho_{\rm B}\omega^2\psi = D\frac{d^4\psi}{dz^4} -T\frac{d^2\psi}{dz^2}.
\end{equation}
We only consider the two extreme cases that the built-in tensile energy is either dominant $T\gg D(L/\delta)^2$ or suppressed $T \ll D(L/\delta)^2$ where $\delta$ is the order of magnitude of the transverse bending.

In this work we consider a three-layer hBN membrane $h\approx 9.5$\AA\ and have used $\rho_{\rm B}=2.1\times10^3$~kgm$^{-3}$, $\sigma=0.211$, $E=850$~GPa~\cite{Falin2017}. A nanoribbon of the length $L=1~\mu$m and the width of $w=3$~nm is considered.

\subsection{Negligible tensile force}
In the limit of negligible tensile force at the clamped points one finds the following normal mode profiles
\begin{align*}
\psi_n(z) = \mathcal{N}_n\Big[&\frac{\cos(\alpha_n\frac{z}{L}) -\cosh(\alpha_n\frac{z}{L})}{\cos(\alpha_n) -\cosh(\alpha_n)} \\
&-\frac{\sin(\alpha_n\frac{z}{L}) -\sinh(\alpha_n\frac{z}{L})}{\sin(\alpha_n) -\sinh(\alpha_n)}\Big],
\end{align*}
where $\mathcal{N}_n$ is the normalization factor which is set such that the maximum of the mode profile equals unity.
Here, $\alpha_n$ is the $n$th root of the transcendental equation $\cos\alpha\cosh\alpha=1$, whose first few values are $\alpha = \{4.730, 7.853,10.996, \cdots\}$ and they assume values closer to the odd-integer multiples of $\pi/2$ as $n$, the number of root, increases.
The normal frequencies are then
\begin{equation}
	\omega_n^2 = \frac{D}{\rho_{\rm B}}(\frac{k_n}{L})^4,
\end{equation}
with the wave number $k_n = \big[\int_0^L dz \psi_n(z)\psi^{(4)}_n(z)\big]^{\frac{1}{4}}$.
The coupling strength of a color center to each of the flexural normal modes depends the local strain imposed on it, which in turn varies with the axial location of the color center
\begin{equation}
\lambda_n(z) = \Xi \varepsilon_n(z) \approx \frac{\Xi}{2L^2}x_{{\rm zp},n}^2[\psi'_n(z)]^2,
\end{equation}
where $\Xi$ is the deformation susceptibility whose value from the DFT analysis is about $3$~PHz and $x_{\mathrm{zp},n} = \sqrt{\hbar/2m_n\omega_n}$ is the zero-point amplitude of the $n$th normal mode whose effective mass is given by $m_n =\rho_{\rm B}hw\int_0^Ldz[\psi_n(z)]^2$, where $w$ is the width of nanoribbon.
To find the wave number of the modes as well as their coupling rates to a color center we employ a numerically tractable approach~\cite{Khasawneh2019}.
\begin{figure}[b]
\includegraphics[width=\columnwidth]{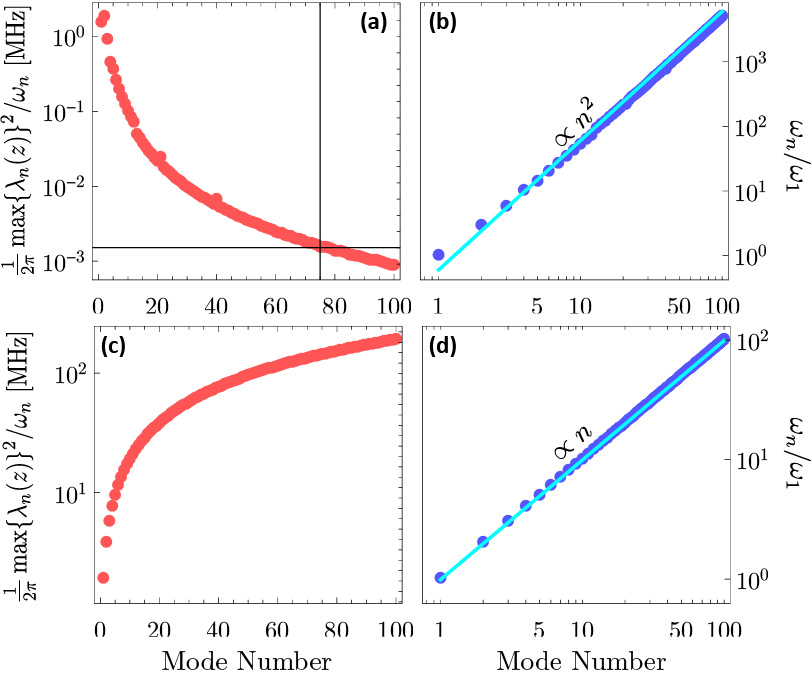}
\caption{%
The maximum contribution of individual flexural modes in the effective coupling between color centers: (a) Clamped and (c) Pinned BNNR.
In (b) and (d) we plot normal mode frequency distributions for the clamped and pinned boundary conditions, respectively.
}
\label{fig:coupling}
\end{figure}

\subsection{Dominant built-in strain}
In the opposite regime where the tensile force at the boundaries are dominant the mode profiles are simply given by
\begin{align*}
\psi_n(z)=\sin(\beta_n \frac{z}{L}),
\end{align*}
where we have introduced $\beta_n \equiv \omega_n\sqrt{\rho_{\rm B}h/T} = n\pi$ with $n$ taking positive integers.
The mode frequencies are then
\begin{equation}
\omega_n = \sqrt{\frac{T}{\rho_{\rm B}h}}\frac{n\pi}{L},~~~(n=1,2,\cdots).
\end{equation}
In this case the effective mass of all modes are equal to $m_n=\half\mu L$. The coupling rates are thus
\begin{equation}
\lambda_n(z)= \frac{n\pi\hbar\Xi}{2L^2\sqrt{\rho_{\rm B}h T}}\cos^2\!(n\pi\frac{z}{L}).
\end{equation}
This equation already shows that the coupling rate linearly increases with the mode number. Therefore, the effective coupling truncation becomes invalid as the higher order modes hold more share.

In Fig.~\ref{fig:coupling} the mode frequency distribution and the effective coupling contribution of the flexural modes in the two above discussed regimes: clamped and pinned boundary conditions, are presented.
We notice that for the case of dominant tensile force at the boundaries the mode frequencies form a commensurate spectrum $\omega_n = n\omega_1$. Nonetheless, the contribution of higher order modes in the qubit-qubit dynamics dominates those of low-frequency modes.
In contrast, in the clamped configuration the higher frequency of the mode, the less it has to do with the coupling of two qubits. However, the mechanical spectrum becomes nonlinear $\omega_n\propto n^2$.
In this work, we have put our focus on the doubly clamped case as it is the case that is usually fabricated experimentally and for its theoretical advantage in the accessibility of a converging coupling spectrum.

\section{Measures of Entanglement}\label{sec:measures}
In this appendix, we bring in the definition of genuine multipartite entanglement and quantum Fisher information.

\subsection{Logarithmic negativity}
For evaluating the amount of entanglement between two color centers we have employed logarithmic negativity as the measure. It is calculated through the singular values that belong to the partial transpose density matrix~\cite{Plenio2005}
\begin{equation}
E_N(\rho_{AB}) = \log_2\Vert\rho_{AB}^{\intercal_A}\Vert_1,
\end{equation}
where $\Vert O\Vert_1$ is the trace norm or sum of the singular values of the operator $O$, and $\rho_{AB}^{\intercal_A}$ denotes the partial transposition with respect to the subsystem $A$.

\subsection{Genuine multipartite entanglement}
We use the genuine multipartite entanglement measure originally introduced in Ref.~\cite{Coffman2000} and later generalized in several works including \cite{Ma2011} and \cite{Szalay2015} to quantify the entanglement in our system.
The measure exploits general bipartite monogamy property of the entanglement in a $N$-party system such that~\cite{Adesso2008}
$$E^{1|2,3,\cdots,N} =\sum_{j=2}^N E^{1|j} +\sum_{k>j}^N\sum_{j=2}^N E^{1|j|k} +\cdots +E^{\underline{1}|2|\cdots|N},$$
where the underline denotes the \textit{focus} party and $E$ is a proper measure of entanglement that guarantees the convexity. Concurrence squared is the proper choice~\cite{Ou2007, Cornelio2013, Szalay2015}.
The genuine residual $N$-partite entanglement is then calculated as the minimum over all permutations of the subsystem indices
\begin{equation}
E^{1|2|\cdots|N} \equiv \min\{E^{\underline{i_1}|i_2|\cdots|i_N}\}.
\end{equation}
\begin{figure}[b]
\includegraphics[width=\columnwidth]{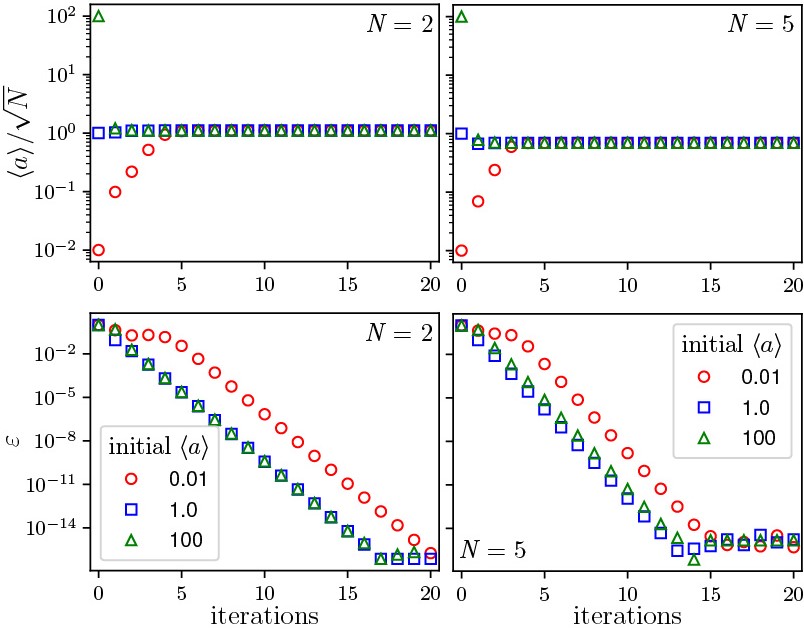}
\caption{%
The values of order parameter $\mean{a}/\sqrt{N}$ (top panels) and the relative error (bottom panels) at each iteration for three different initial guesses. The left panels correspond to a system with two color centers, while the right panels are presenting the results for a system composed of five qubits.
}
\label{fig:method}
\end{figure}

\subsection{Quantum Fisher information}
The quantum Fisher information determines the Cramer-Rao bound in parameter estimation and saturates to the Heisenberg limit for a fully entangled system~\cite{Giovannetti2011}.
Hence, it reflects the degree of multipartite entanglement~\cite{Hyllus2012}.
For a mixed state $\rho$ and observable $O$ the QFI is defined as
\begin{equation}
F_Q[\rho,O]=2\sum_{k,l}|\bra{k}O\ket{l}|^2\frac{(\lambda_k-\lambda_l)^2}{\lambda_k+\lambda_l},
\end{equation}
where $\lambda_k$ and $\ket{k}$ are the eigenvalues and eigenvectors of $\rho$, respectively.
The sum is over indices that $\lambda_k+\lambda_l>0$~\cite{Braunstein1994}.
In this work, we take the collective spin operators $J_n \equiv \vec{n}\cdot \vec{J} = \sum_{\alpha=x,y,z}n_\alpha J_\alpha$ as the observable.
Here, $J_\alpha \equiv \frac{1}{2}\sum_{k=1}^N \sigma_k^\alpha$ is the collective spin operator and $\vec{n}$ is a unit vector that determines components of each spin direction.
 and introduce $\bar{F}_{\!J_n}\equiv \frac{1}{N}F_Q[\rho,J_n]$, the normalized QFI.
This quantity is then upperbounded by $N$ for a fully entangled system.

\section{The iterative method}\label{sec:method}
In this appendix we show the convergence of our numerical method for studying the phase transition in the Dicke-Ising model.
In Fig.~\ref{fig:method} a typical convergence tracking of the iterative result for generating each data point in Figs.~\ref{fig:dicke} is given.
We observe that the method is very robust against the initial guess for $\mean{a}$ and converges very rapidly. This is clear from the relative error at each iteration which is defined as $\varepsilon_i\equiv \big|\abs{\mean{a}_{i}}-\abs{\mean{a}_{i-1}}\big|/(\abs{\mean{a}_{i}}+\abs{\mean{a}_{i-1}})$, where $\mean{a}_i$ is the value of $\mean{a}$ at the $i$th iteration.

%
%
\bibliography{strain}

\end{document}